\documentclass[aps,prl,preprint,nofootinbib,floats,superscriptaddress]{revtex4}
\usepackage{amssymb,graphics,graphicx, epsfig}
\usepackage{graphicx}
\usepackage{amsfonts}
\usepackage{amssymb}
\usepackage{bm}

\newcommand{\beqn}{\begin{eqnarray}}
\newcommand{\eeqn}{\end{eqnarray}}
\newcommand{\be}{\begin{equation}}
\newcommand{\ee}{\end{equation}}

\newcommand{\sm}{{standard ~model~}}
\newcommand{\st}{{Stueckelberg~}}

\newcommand{\mathsym}[1]{{}}

\begin{document}

\title{PAMELA Positron Excess  as  a Signal from the Hidden Sector}
\author{Daniel Feldman$^1$, Zuowei Liu$^{2}$, and Pran Nath}
\affiliation{Department of Physics, Northeastern University,
 Boston, MA 02115, USA,\\
 $^2$C. N. Yang Institute for Theoretical Physics, Stony Brook University, Stony Brook, NY 11794, USA.\\ {\rm V1 Dated: Oct 31, 2008.}}
\begin{abstract}

The recent positron excess observed in the PAMELA satellite experiment
 strengthens previous
experimental findings. We give here an analysis of this excess in the
framework of the Stueckelberg extension of the standard model 
which includes an extra $U(1)_X$ gauge field and  matter in the hidden sector.
Such matter can produce the right amount of dark matter consistent with
the WMAP constraints. Assuming the hidden sector matter to be  Dirac fermions it is shown that 
their annihilation can produce
the positron excess  with the right positron energy dependence 
seen in the HEAT, AMS  and the PAMELA  experiments. The predictions of the $\bar p/p$ flux ratio
also fit the data. 
\begin{flushleft}

\end{flushleft}
 \end{abstract}
 \maketitle
 \preprint{NUB-3262 }
 \preprint{YITP-SB-08-50}
 
{\em Introduction:} An excess of the positron flux emanating from  the galactic halo has been reported by 
the PAMELA satellite experiment\cite{Adriani:2008zr} which supports the previous observations by the 
HEAT and AMS experiments \cite{Barwick:1997igAguilar:2007yf}   but is much more accurate.
A remarkable feature of the positron spectrum is the turn around and increase in the positron flux with 
positron energy in the range of 10-80 GeV.   
Here we analyze the possibility that the positron flux is arising from the annihilation of  particles in the 
hidden sector. The hidden sector, which is defined as  a sector with fields which are neutral under
the standard model (SM)  gauge group, 
 has played an increasingly important role since its inception in the formulation of 
supergravity grand unification\cite{Chamseddine:1982jx}.
 Subsequently, 
 it was realized that such hidden sectors appear quite naturally in the context of string 
 theory\cite{Candelas:1985en}. 
  More recently the hidden sector was utilized in the Stueckelberg extension of the standard model.
 In such an extension, the SM gauge group $SU(3)_C\times SU(2)_L\times U(1)_Y$  is supplemented by 
 an extra $U(1)_X$ gauge group factor\cite{kn,fln1}  where the SM fields are neutral under the $U(1)_X$.
 However, the mixings  between the SM and the hidden sector do occur via a connector sector
 which mixes the gauge fields of $U(1)_Y$ and of $U(1)_X$. Such mixings can occur via the \st 
 mass terms\cite{kn} or via kinetic mixings\cite{holdom} or both\cite{Feldman:2007wj}.
 Electroweak constraints from the LEP and the Tevatron were analyzed in \cite{kn,fln1,Feldman:2007wj}
and an analysis of the dark matter was given with and without kinetic mixing in  
   \cite{Feldman:2006wd,Cheung:2007ut,Feldman:2007wj} consistent with the
WMAP\cite{Spergel:2006hy} constraints.
(For related works on the $U(1)$ extensions see\cite{holdom,Kumar:2006gmChang:2006fp}, 
 and for other works on the hidden sector see \cite{HSmodels}). 
Further recent works regarding \st  extensions in the  context
 of the string and D brane models can  be found in \cite{CorianoAnas} 
and for related works see \cite{KumarAbelBurgess}. 
We note that hidden sectors are also central to unparticle\cite{Georgi:2007ek} and ungravity models \cite{Goldberg:2008zz}.

  {\em Positron fraction from annihilation in the hidden sector:} 
The analysis of the positron spectrum depends both on the particle physics as  well as  on the  astrophysical
  models and these features have been discussed recently in some detail in \cite{Grajek:2008jbBarger:2008su,Chen:2008yi}.  
  Here we focus on the fit to the just released  data by the PAMELA experiment\cite{Adriani:2008zr} from 
  annihilation of dark matter in the hidden sector in the framework of \st extension of the standard model.
 We give now the details of the analysis.

In general the positron flux arising from the annihilation of dark matter (DM) particles 
  is given by 
 \cite{Delahaye:2007fr,Cirelli:2008id} 
\be
\Phi_{e^{+}} = \frac{\eta B  v_{e^{+}} }{4 \pi b(E)}\frac {\rho^2_{\rm }}{M^2_{D}} \int^{  M_{D}  }_{  E }  \sum_k{\langle \sigma v \rangle_{kH}\left(\frac{d N_{e^{+}}}{d E'}\right)_k } {\cal I }_{(E,E')}d E'
\label{1}
\ee
where $M_D$ is the mass of the dark matter particle, 
 $\eta =1/2(1/4)$ for the DM particle being Majorana or  Dirac\cite{Delahaye:2007fr},
 B is the boost factor which is expected to lie in the range (2-10) 
although significantly larger values have been used in the literature. In the above  $v_{e^{+}}$ is the positron velocity where
  $v_{e^{+}} \sim c$, and $\rho$ is the local dark matter density in the halo so that $\rho$ 
  lies in the range
   $(0.2-0.7) [\rm GeV/cm^3]$ \cite{Kamionkowski:2008vw}. 
Further,  $b(E)$  in Eq.(\ref{1}) is  given by \cite{Longair:1994wu,Hooper:2004bq,Hisano:2005ec} 
$b(E)  = E_0 (E/E_0)^2 /\tau_E$, where  $\tau_E  \sim 10^{16} [\rm s]$, with $E$ in [GeV] and $E_0 \equiv 1 \rm GeV$.
Here $\langle \sigma v \rangle_{H} $ is the velocity averaged cross section
in the {\em Halo (H) of the galaxy} as emphasized by the subscript {\em H}. 
In some works 
 $\langle \sigma v \rangle_{H}$
is  replaced  by the $\langle \sigma v \rangle_{X_f}$ at the freezeout temperature. However, such an
approximation can lead to significant 
errors since the ratio $\langle \sigma v \rangle_{H}/\langle \sigma v \rangle_{X_f}$ can deviate significantly
from unity depending on the part of the parameter space one is in.
The halo function ${\cal I }_{(E,E')}$ 
is parametrized as in \cite{Cirelli:2008id}, and we consider both the Navarro, Frenk and White (NFW) and Moore
et. al \cite{Navarro:1996gjMoore:1999gc} profiles  coupled with various diffusion models. 

\begin{figure*}[t]
\includegraphics[width=8cm, height=6cm]{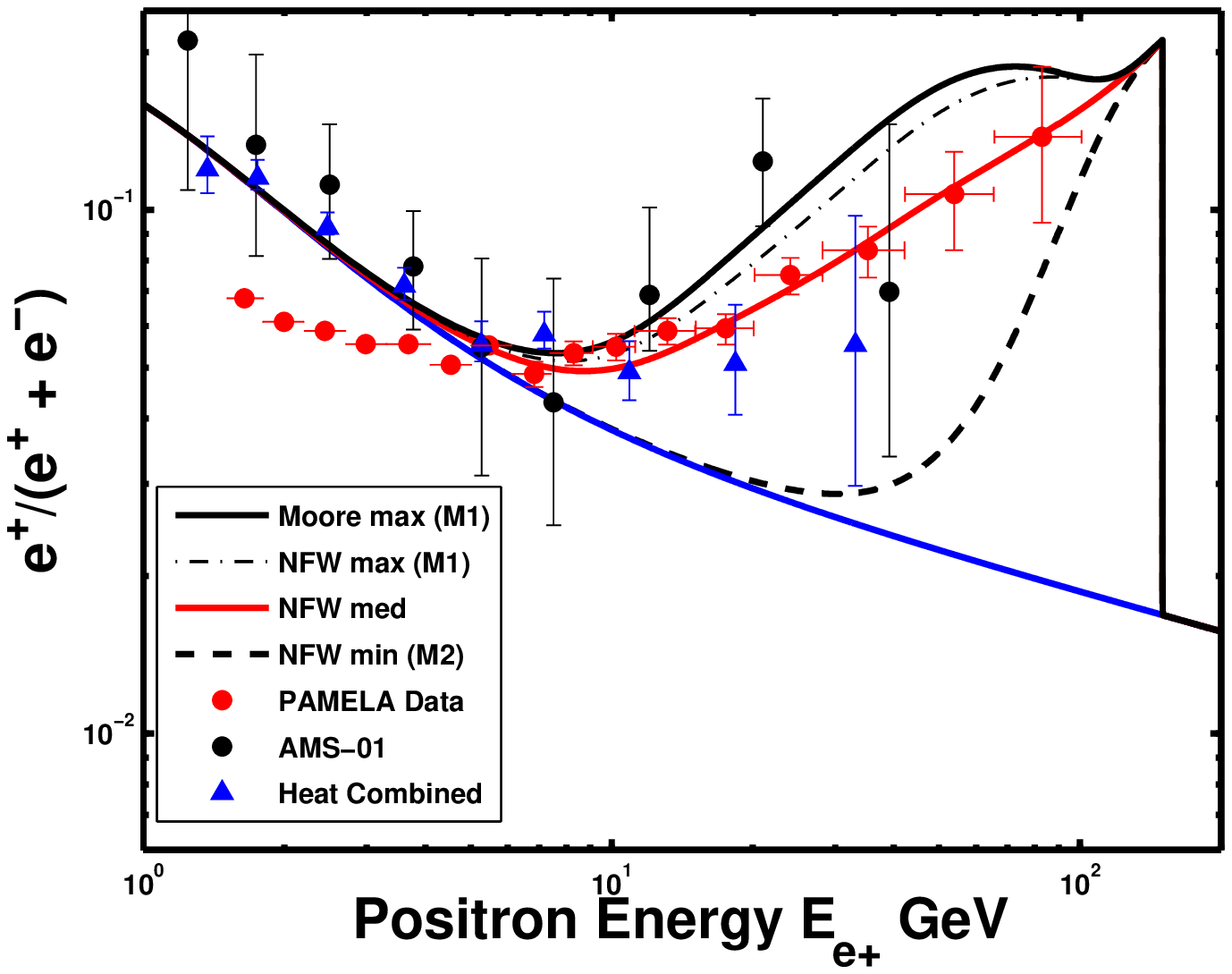}
\includegraphics[width=8cm, height=6cm]{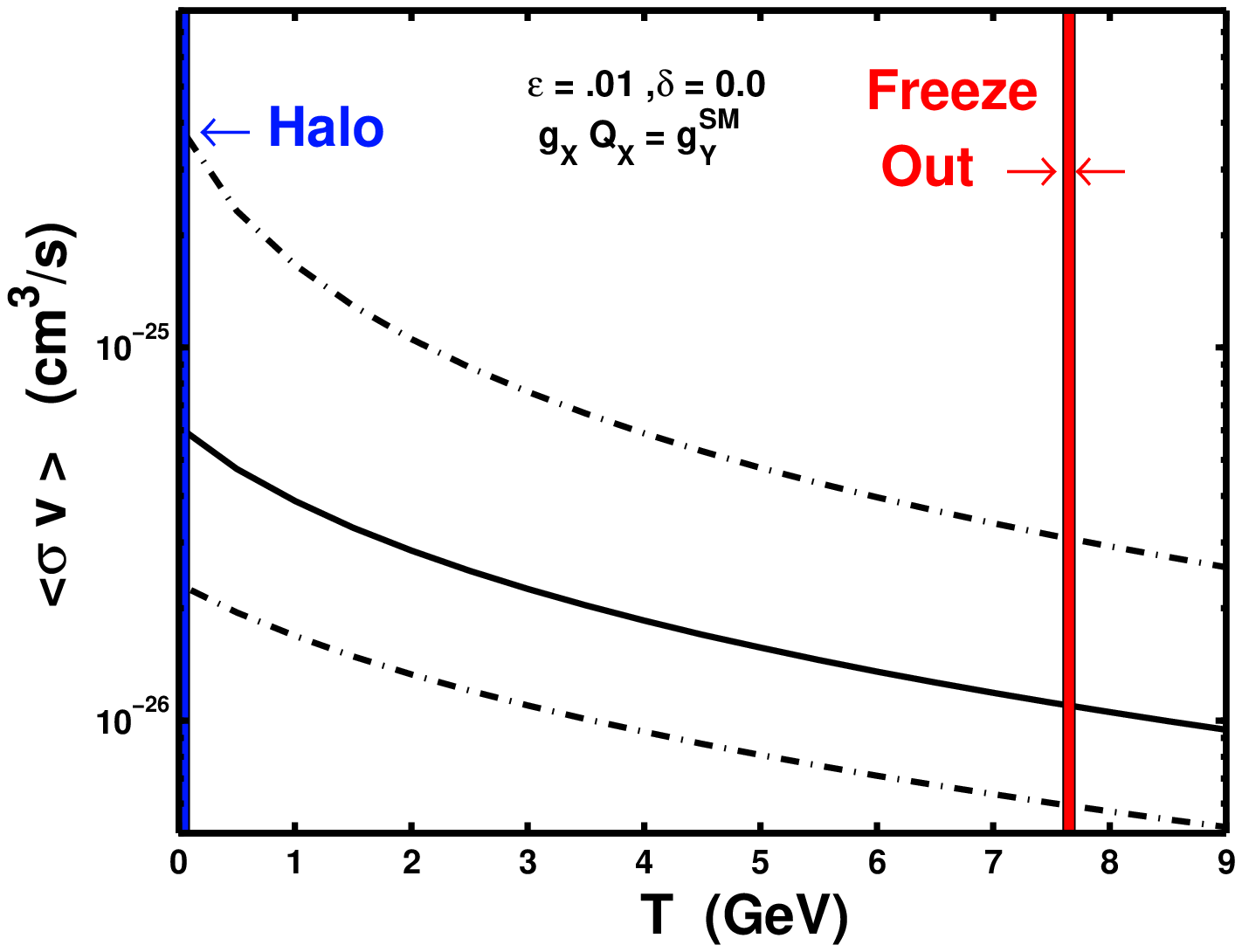}
\caption{ (Left Panel): Positron spectrum including the monochromatic source 
and continuum  flux for various halo/diffusion models with $(\epsilon= 0.006,\delta =0.00)$ and 
  $\rho= 0.35~\rm GeV/cm^3$  with $M_{Z'}= 298$ GeV, 
$M_D=150$ GeV, $\tau_E = 3 \times 10^{16} \rm s$\cite{Longair:1994wu}, and  B=10; and $\Omega h^2 = 0.13$
(calculated by integration over the Breit-Wigner pole).
 Also plotted is
 the just released PAMELA  data \cite{Adriani:2008zr},
 along with the AMS-01 and HEAT data \cite{Barwick:1997igAguilar:2007yf}.  The background flux ratio is the decaying solid (blue) lower curve. (Right Panel):
An exhibition of the  dependence of 
$\langle \sigma v\rangle$ on temperature for  \st models  as given
in the figure with  $M_D/\rm GeV \in [150,153]$ in steps of 1.5
and $M_{Z'}$ fixed as in the left panel of  Fig(\ref{1}).
The annihilation near a pole generates a significant enhancement of $\langle \sigma v\rangle_{H}$
in the halo relative to $\langle \sigma v\rangle_{X_f}$ at freezeout. The natural Breit-Wigner enhancement of 
$\langle \sigma v\rangle_H$ obviates the necessity of using very large boost factors.
}
\label{fig1}
\end{figure*}


The  channels that can contribute to the positron flux in 
the model are $D\bar D \to f \bar f, W^+W^-,\ldots$
where $f$ is any quark or lepton final state of the standard model.
In our analysis we make use of the dominance of the direct channel 
poles,  specifically of  the $Z'$ pole,  and here the $W^+W^-$ final state
contribution is much suppressed relative to the $f\bar f$ final state contribution\cite{fln1}.
We begin by discussing the $D\bar D\to f\bar f$ cross section in  \st extension of  the \sm.
Here one has a mixed Higgs mechanism and a \st mechanism to break the extended
 $SU(2)_L\times U(1)_Y\times U(1)_X$  electroweak symmetry \cite{kn}. 
 After such breaking there is the usual mixing between the neutral component of the $SU(2)_L$
 gauge field $A_3^{\mu}$ and the $U(1)_Y$ gauge field $B^{\mu}$. Thus together the Higgs 
 mechanism and the \st mechanism generate a $3\times 3$ mass matrix for the three gauge fields
 $C^{\mu}, B^{\mu}, A_3^{\mu}$.  
The above model has a  non-diagonal mass
matrix  (${M}^2_{\rm St}$)\cite{kn} and a 
 non-diagonal kinetic mixing matrix  ($\mathcal{K}$)\cite{holdom,Feldman:2007wj}.
A diagonalization of the kinetic  term can be obtained by  a $\rm GL(3)$  transformation. 
However, such a diagonalization is arbitrary up to an orthogonal transformation. One can choose a combination
$K'$ of $\rm GL(3)$ and an orthogonal transformation such that ${\cal{M}}^2= K^{'T} M^2_{St} K^{'}$ has a form
similar to that of $M^2_{St}$, i.e., the zero elements of $M^2_{St}$ are maintained by the transformation
(for details see \cite{Feldman:2007wj}).
In the basis where both the kinetic  and the mass$^2$ matrix are
diagonal one has
$ \mathcal{M}_{Diag}^2 ={\mathcal{ R}}^T\mathcal{M}^2 {\mathcal{ R}}$.
In this basis the interactions between the standard model fermions and the  vector bosons
$Z, Z'$  are given by
$ {\cal L}^{\rm VS}_{\rm int} =  \bar f \gamma^\mu \left[
  C^{Z'}_{f_{L}} P_L Z'_\mu
 + C^{Z}_{f_{L}} P_L Z_\mu + ( L\to R)\right]f$,
where as usual  $Q = T^3_{L,R}+Y_{L,R}/2$, $T^3_{R} = 0$,  $P_{L,R}=\frac{1}{2}(1 \mp \gamma^5)$,
and 
$C_{f_{L}}^{Z_{i}} = T^3_{L}\left[ g_2 {\mathcal{ R}}_{3i} - \gamma \sqrt{1+\bar \epsilon^2}{\mathcal{ R}}_{2i}\right]$,
 $C_{f_{R}}^{Z_i} = Q\gamma \sqrt{1+\bar \epsilon^2}{\mathcal{R}}_{2i}$,   where
$Z_i= (Z', Z)$ for $i=1,2$.
 The photon couples as usual to the visible fermionic fields with strength $e Q$, and 
 $\gamma$ is related to $e$ by $e=\gamma g_2/[\gamma^2+g_2^2]^{1/2}$.
 The Dirac  fermion is assumed to have the interaction $g_XQ_X \bar D\gamma^{\mu} D C_{\mu}$ which
 in the diagonal  basis gives 
 $ {\cal L}^{\rm HS}_{\rm int} =   \bar D \gamma^\mu \left[
C^{Z'}_D  Z'_\mu+ C^Z_D  Z_\mu +C^\gamma_D A^{\gamma}_\mu \right]D$, 
 where 
 $C^\gamma_D = g_X Q_{X} (- c_\theta s_\phi -S_{\delta} c_\theta c_\phi)$, 
   $C^Z_D = g_X Q_{X} (s_\psi c_\phi + s_\theta s_\phi c_\psi -S_{\delta}(s_\psi s_\phi - s_\theta c_\phi c_\psi))$, 
  $C^{Z'}_D = g_X Q_{X} (c_\psi c_\phi - s_\theta s_\phi s_\psi-S_{\delta}(c_\psi s_\phi + s_\theta c_\phi s_\psi))$.
Here $S_{\delta} =\delta/\sqrt{1-\delta^2}$ and the angles $\theta, \phi, \psi$ appear in the 
 rotation matrix ${\mathcal{R}}$,  and are defined by
 $\tan \theta = \gamma/g_2$,  $\tan \phi = \bar \epsilon = [\epsilon -\delta][1-\delta^2]^{-\frac{1}{2}}$ and  $\psi$ is determined 
by the relation 
$\tan 2\psi= { 2\bar{\epsilon}  M_{0}\sqrt{M^2_0-M^2_W}}/(
{M^2_1-M^2_0+\bar{\epsilon}^2(M^2_0 + M^2_1-M^2_W)})$,
where  $M_0= v\sqrt{g_2^2 +\gamma^2}/2$.
The parameters $\epsilon$ 
and $\delta$ are constrained by the electroweak data\cite{fln1,Feldman:2007wj}. 
One finds that  $\epsilon$ and $\delta$ are both separately constrained so that
$|\epsilon |, |\delta | \lesssim .06$. 
The action given in \cite{kn} leads to an integrated cross section\cite{Cheung:2007ut},
\beqn
\sigma_{f \bar f} \simeq \frac{N_f s}{32\pi} \frac{\beta_f}{\beta_D} [( |\xi_{L}|^2 +|\xi_{R}|^2)\cdot F_{1} + Re(\xi_L^*\xi_R)\cdot F_{2}],
\eeqn
where 
$F_{1}=    1+  \beta_D^2 \beta_f^2/3  
+{4 M_D^2}{s}^{-1} \left(1-{2m_f^2}/{s}\right )$, and
$F_{2}  =   8  {m^2_f}{s}^{-1} \left(1+{2M_D^2}/{s}\right )$.
Here $\beta_{f,D}=(1-4m^2_{f,D}/s)^{1/2}$,  
 $s=4m^2_{D}/(1-v^2/4)$ and $\xi_{L,R}$  include the poles 
\be
  \xi_{L,R} = \frac{C^\gamma_{D} e Q }{s}
           +\frac{C^Z_{D} C_{f_{L,R}}^Z }{s- M_Z^2 + i \Gamma_Z M_Z}
           +\frac{C^{Z'}_{D} C_{f_{L,R}}^{Z'} }{s - M_{Z'}^2 + i \Gamma_{Z'} M_{Z'}}.
\label{7}
\ee
The  dominant term in our analysis is the line source arising from  the annihilation $D\bar D \to Z'\to e^+e^-$
and in this case one has 
$
\sum_{ F =  \rm Final~states} \langle \sigma v \rangle_{F} \left({d N_{e^{+}}}/{d E'}\right)_F  
\sim \langle \sigma v \rangle_{e^{+} e^{-}} \delta(E'-M_{D})  + \ldots
$,
where the dots stand for the background terms that contribute to the continuum  flux. 
The continuum  flux arises mostly from muons and to a much lesser degree from taus\cite{Cirelli:2008pk}.  
 Defining $R_f$ as 
the positron ratio from source $f$  
one finds $R_{\mu}/R_{e}$ is non-negligible
and decreases with increasing $E_{e^{+}}$ over the dark matter mass
(DM) mass range of interest and a similar relation holds for the taus\cite{Cirelli:2008pk}.  
The inclusion of the flux from the continuum
reduces the needed boost factor slightly, however the line source still dominates at high energies.
The use of the above  
in Eq.(\ref{1}) yields the primary positron flux $\Phi_{e^{+}} \equiv \Phi^{(1)}_{e^{+}}$. 
We must add to it the secondary positron flux 
$\Phi^{(2)}_{e^{+}}$ and  then compare it with the electron flux $(\Phi^{(1)}_{e^{-}}+\Phi^{(2)}_{e^{-}})$. 
For $\Phi^{(1)}_{e^{-}}, \Phi^{(2)}_{e^{-}},\Phi^{(2)}_{e^{+}}$ we use the parametrizations of
\cite{Moskalenko:1997gh,Baltz:1998xv}.
For comparison with experiment one often defines the positron fraction
 $e^+/(e^++e^-)$, and an analysis is given of this observable as a function of the positron energy 
 in Fig.(\ref{fig1}) for the Stueckelberg $Z'$ model. One finds that  the annihilation of Dirac fermions 
 via the $Z'$ pole  into $e^+e^-+ \mu^{+} \mu^{-}$ gives a sufficient kick to generate the necessary turn around in 
 the positron fraction at just about the desired value of the positron energy consistent with the relic density 
 constraints. 
 The analysis of Fig.(\ref{fig1}) (left panel) exhibits the theoretical evaluation for 
 several model points.  Here we consider NFW min (M2), med and max (M1) as well as
the Moore max (M1) parametrizations \cite{Delahaye:2007fr,Cirelli:2008id}.
One finds that there is a significant variation in the prediction depending on the profile/diffusion model one
 chooses.  However, one finds that the PAMELA data does lie in the 
 range of the  theoretical predictions. We note in passing that the gamma ray spectrum in this model has been discussed in\cite{Cheung:2007ut}.
 The theoretical predictions cover a range which includes the PAMELA data\cite{Adriani:2008zr}. 
  Further, such a fit 
 determines the dark matter fermion mass  to be  roughly half  the $Z'$ mass.
 
\begin{figure}[t]
\includegraphics[width=8cm, height=6cm]{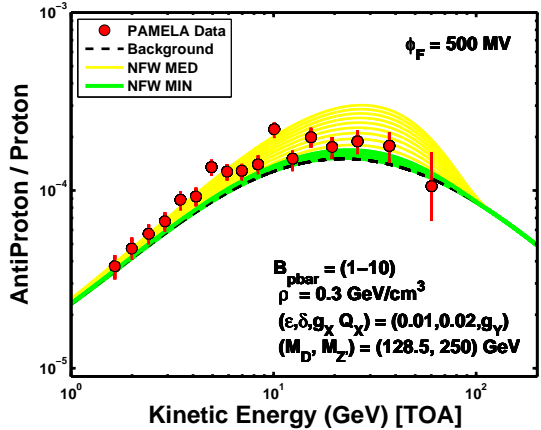}
\caption{The $\bar p/p$ flux ratio including
the TOA correction to the IS spectrum \cite{Bottino:1994xs}, and with $B_{\bar p} \in (1-10)$. The green (darker) curves (NFW min)
are insensitive to the boost in the ratio, while the yellow (lighter) curves (NFW med) allow a boost
as large as 5 or even larger.
}
\label{fig4}
\end{figure}
 
 In Fig.(\ref{fig1}) (right panel)  we exhibit the  dependence of $\langle \sigma v\rangle$ on temperature. The analysis shows that $\langle \sigma v\rangle$
  can have a significant temperature dependence.  Thus the simplifying
 assumption often made in assuming that  $\langle \sigma v\rangle$ is a constant as one moves from
 the freezeout  temperature to the temperature of the galactic halo is erroneous.  Specifically the
 analysis shows that the temperature dependence is model dependent and 
 one can generate an enhancement of  $\langle \sigma v\rangle_{H}$ in the halo relative to 
freezeout  $\langle \sigma v\rangle_{X_f}$  by as much as a factor of 10 or more depending on
 the part of the parameter space one is in. Typically the temperature dependence  is
 enhanced when the dark matter particles  annihilate near a pole from the Breit-Wigner which is the case in the analysis
 here. 
 
\begin{figure}[htb]
\includegraphics[width=7.5cm, height=5.5cm]{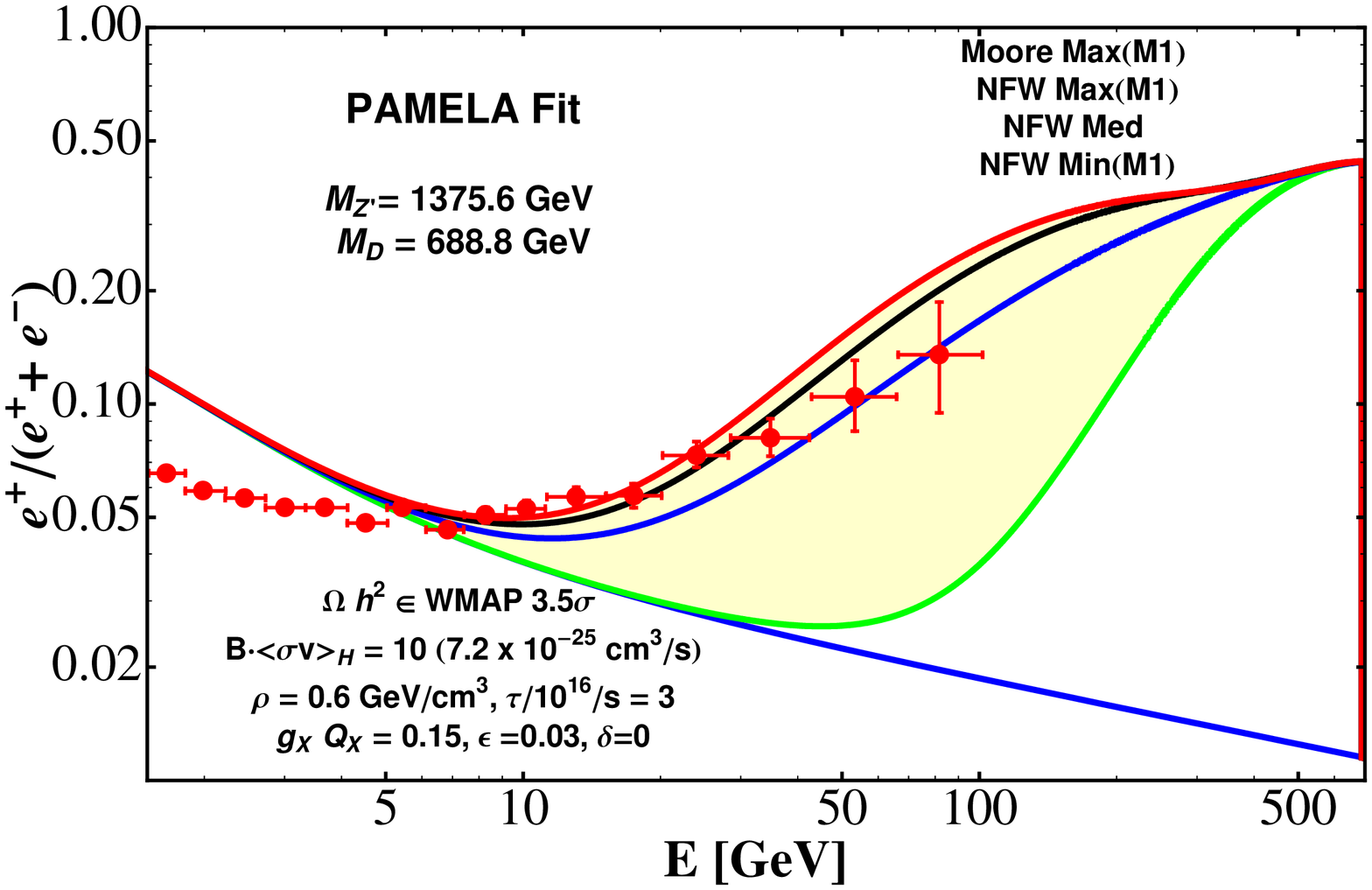}
\includegraphics[width=7.5cm, height=5.5cm]{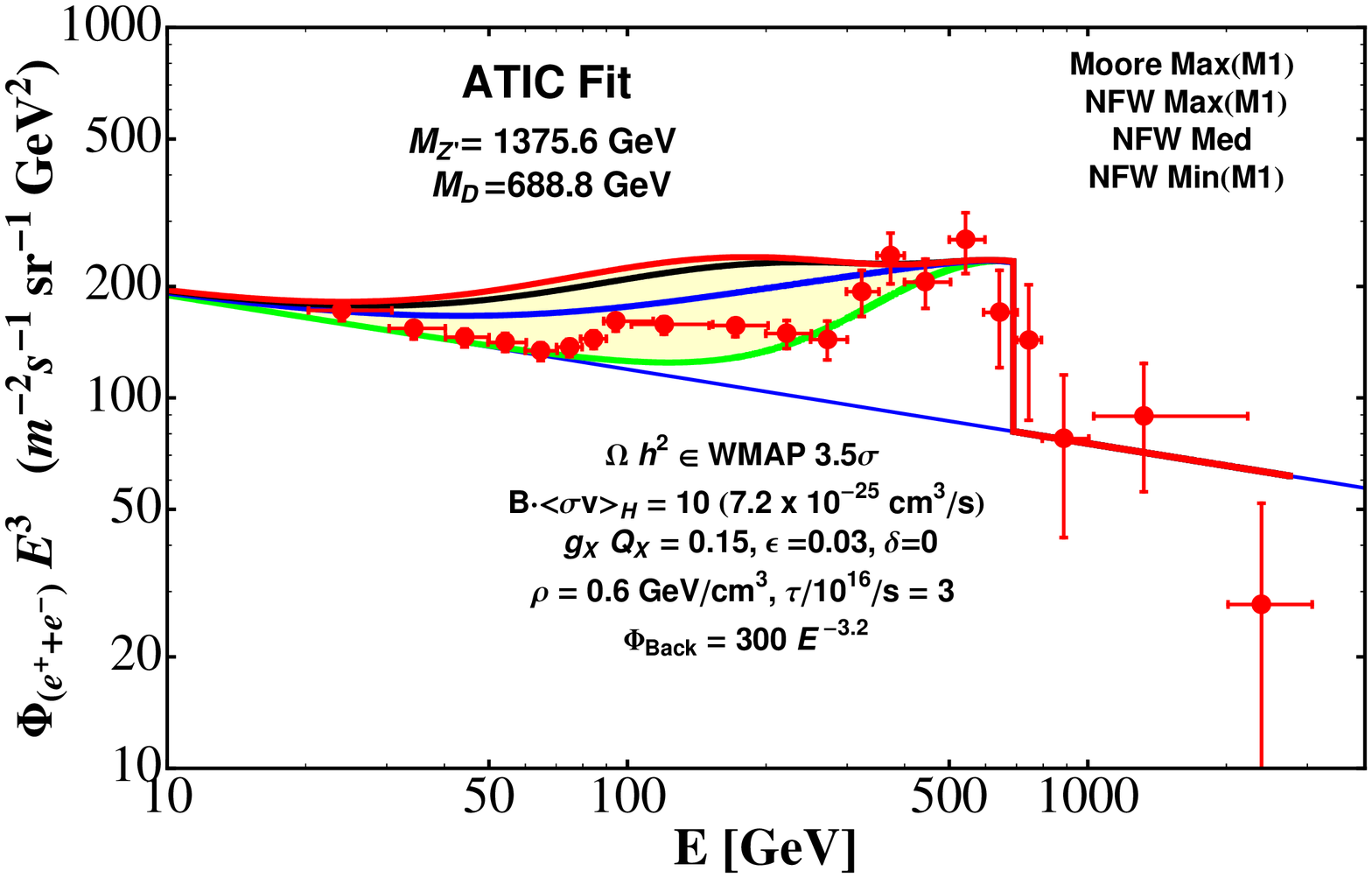}
\caption{
Fit to the PAMELA and ATIC\cite{:2008zzr} data for a heavy
Dirac dark matter mass of $688.8$ GeV with the Breit-Wigner 
enhancement.  The  curves in descending order are for the cases 
for the halo profiles listed on the top right hand corner. 
}
\label{fig5}
\end{figure}

{\em Anti-proton Flux and PAMELA :}
The $\bar p/p$ flux ratio as recently reported by the PAMELA\cite{Adriani:2008zq} collaboration
  indicates a smooth increase  with energy
up to about  10 GeV and then a flattening  out    in agreement with the background
and with previous experiments.
We note that a suppression of $\bar p/p$ flux ratio is possible in the model presented here. This is due in part 
  because
the $Z'\to W^+W^-$ is suppressed  as already discussed. We have carried out a detailed
analysis of the  $\bar p/p$ flux ratio.
Our analysis follows closely the work of \cite{Cirelli:2008id}
with fragmentation functions as modeled in Bottino et al and by Bergstrom etal 
and  ($p,\bar p$) backgrounds as in Donato et al and Bringmann et al \cite{Bottino:1994xs}.
The Interstellar (IS) flux has been modified for predictions at the Top of the Atmosphere (TOA)
which suffers from large uncertainties. 
The results are given in Fig.(\ref{fig4}) and compared with the 
recently reported results by the PAMELA collaboration. 
It is found that the $\bar p/p$ analysis of Fig.(\ref{fig4}) is fully
compatible with the recent PAMELA data.
It is further observed
that the NFW min profile, for the $\bar p/p$  predictions,
are rather insensitive to a boost factor, while boost factors
as large as 5 or larger are acceptable in the NFW med model.   We note in passing
that the  $\bar p/p$ flux ratio does suffer from larger theoretical uncertainties
than the $e^{+}/e$ flux ratio due to a larger diffusion length. 
Further, it is known 
that local inhomogeneities in the dark matter density 
may lead to very different boost factors for positrons
and antiprotons (see, for example,  Lavalle etal in  \cite{Bottino:1994xs}).

{\em Conclusion:}
In this work 
we have shown that the annihilation of the Dirac fermions in the hidden sector close to the $Z'$ pole can generate 
a  positron fraction compatible with the current PAMELA data. 
Specifically the model produces  
the right amount of positron spectrum enhancement with increasing positron energy indicated by the
 AMS-01 and the HEAT data and confirmed by the PAMELA data, and additionally the model can  accommodate the antiproton constraints.
 A further support of the model can come from a direct  observation of the $Z'$ boson at the Large Hadron Collider.
 
 {\em Note Added:}
After submission of this paper, the ATIC Collaboration published its data on the electron excess\cite{:2008zzr}.
One can fit both PAMELA and ATIC as well as the $\bar p/p$ flux 
in the model with a change of the dark matter and $Z'$ mass and  with
a low boost factor of 10. This fit is given in Fig.(\ref{fig5}) and includes the continuum  flux. Again the Breit-Wigner 
enhancement plays an important role in the analysis. 

{\em Acknowledgments}:  
This research is  supported in part by NSF grants PHY-0653342 (Stony Brook)  and PHY-0757959 (NU). 

\vspace{-.5cm}

\end{document}